\documentclass[final,3p,times]{elsarticle}
\usepackage{amssymb}
\usepackage{amsmath}
\usepackage{color}
\begin{document}
%% Definitions
\def\nuc#1#2{${}^{#1}$#2}
\def\mnu{$\langle m_{\nu} \rangle$}
\def\mb{$\langle m_{\beta} \rangle$}
\def\BBz{0$\nu\beta\beta$}
\def\BBm{$\beta\beta(0\nu,\chi)$}
\def\BBt{2$\nu\beta\beta$}
\def\BB{$\beta\beta$}
\def\Gz{$G^{0\nu}$}
\def\Mz{$|M_{0\nu}|$}
\def\Mt{$|M_{2\nu}|$}
\def\Tz{$T^{0\nu}_{1/2}$}
\def\Tt{$T^{2\nu}_{1/2}$}
\def\mj{M{\sc ajo\-ra\-na}}
\def\QBB{Q$_{\beta\beta}$}

\begin{frontmatter}

%% Title, authors and addresses
\title{Systematic Effects in Pulse Shape Analysis of HPGe Detector Signals for \BBz}
\author[label1,label2]{V.M. Gehman}\ead{vmg@lanl.gov}
\author[label1]{	S.R. Elliott	}
\author[label3]{D.-M. Mei}
\address[label1]{Los Alamos National Laboratory, Los Alamos, NM 87545}
\address[label2]{Center for Experimental Nuclear Physics and Astrophysics, and Department of Physics, University of Washington, Seattle, WA 98195}
\address[label3]{Department of Physics, The University of South Dakota, Vermillion, SD 57069}

%% Text of abstract
\begin{abstract}
Pulse shape analysis is an important background reduction and signal identification technique for next generation of \nuc{76}{Ge} \BBz\ experiments.  We present a study of the systematic uncertainties in one such parametric pulse-shape analysis technique for separating multi-site background from single-site signal events.  We examined systematic uncertainties for events in full-energy gamma peaks (predominantly multi-site), double-escape peaks (predominantly single-site) and the Compton continuum near \QBB\ (which will be the dominant background for most \BBz\ searches).  In short, we find total (statistical plus systematic) {\it fractional} uncertainties in the pulse shape cut survival probabilities of: 6.6\%, 1.5\% and 3.8\% for double-escape, continuum and $\gamma$-ray events respectively.
\end{abstract}


\end{frontmatter}

%% \linenumbers

%% main text
\section{\label{sec:Intro}\BBz\ in Germanium}
The search for physics beyond the Standard Model has one of its most promising leads in neutrinoless double-beta decay (\BBz) in particular, and neutrino physics more generally.  Interest in \BBz\ is extremely well-motivated in the literature \cite{Elliott02, Elliott04, Avignone05, Barabash04, Ejiri05, Avignone07}.  There are approximately 10 isotopes known to undergo two-neutrino double beta decay (\BBt, {\it i.e.} $^{76}\mbox{Ge} \rightarrow ^{76}\mbox{Se} + 2 e^{-} + 2\overline{\nu}$), the slowest nuclear decay allowed in the Standard Model.  These are also the isotopes of interest in the search for \BBz\ ({\it i.e.} $^{76}\mbox{Ge} \rightarrow ^{76}\mbox{Se} + 2 e^{-} + 0\overline{\nu}$), a \BB\ mode forbidden by the Standard Model.  If observed, \BBz\ would imply the existence of massive Majorana neutrinos\cite{Schechter82} and could also lead to the discovery of other physics beyond the Standard Model.  If \Tz\ could be measured in several isotopes, it would do much to elucidate this physics \cite{Gehman07}.  To make a meaningful comparison between \Tz\ measurements in several isotopes, the total experimental uncertainty must be sufficiently low.  This places significant demands on the size of systematic uncertainties of the global experimental program.  The importance of systematic uncertainties in potential results of \BBz\ searches motivates the investigation of systematic uncertainties in pulse shape analysis (PSA) presented here.

\nuc{76}{Ge} is one of the \BB\ isotopes under investigation in the current generation of \BBz\ searches (the most stringent \Tz\ limits already come from \nuc{76}{Ge}), and there are two next-generation \nuc{76}{Ge} experiments currently under development: \mj \cite{MJProposal06, Avignone07TAUP, CISNP08, IEEENSS08, MAJORANA_CIPANP} and GERDA \cite{GERDA}.  This article will present work performed in support of the \mj\ experiment.  Both \mj\ and GERDA, as well as all \BBz\ searches (and indeed all searches for rare events), rely heavily on reducing backgrounds while retaining signals as effectively as possible.  One of the ways that both experiments plan to reduce background and identify signal events is through PSA on the array of high-purity germanium (HPGe) detectors that will comprise the experiments.  Because of the inherent complexity of making pulse shape cuts, systematic uncertainties in PSA-cut efficacy could be one of the leading contributors to the total systematic uncertainty budget of these experiments.  An examination of these systematic uncertainties is the primary focus of this article.  We will begin with a description of the detector used in this work, then move on to an overview of the use of PSA in HPGe detectors for \BBz\ searches up to and including this article.  We will then detail our survey of systematic effects on PSA, and close with some discussion of the context of this work in the \mj\ experiment.

\section{\label{sec:Hardware}Description of Experimental Apparatus}
\begin{figure}[htbp]
\centering
\includegraphics[angle=0, width=5cm]{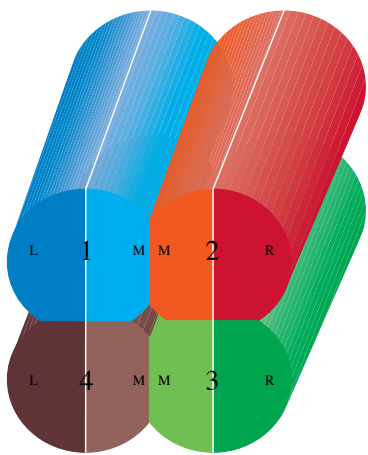}
\caption{A cartoon of the configuration and readout channels of the four detectors in the CLOVER detector used in this study. The central contact of each detector is numbered one through four, and left two, middle four, and right two segmentation contacts are labeled with an ``L,'' ``M,'' and ``R'' respectively.  This Figure is adopted from Reference \cite{Canberra}.}
\label{fig:CloverCartoon}
\end{figure}

The data presented in this article were taken using a CLOVER detector at Los Alamos National Laboratory.  A CLOVER is a commercially available detector system from Canberra \cite{Canberra}, consisting of four, n-type detectors in a single cryostat.  The detectors in the CLOVER used in this study had two-fold azimuthal segmentation.  These characteristics make the CLOVER a good off-the-shelf test bed for potential analysis techniques for \mj.  The individual detectors have a relative efficiency of 26\%, this corresponding to a mass of roughly 750 g per detector.  The CLOVER is instrumented with four high-resolution, cold-FET energy readouts (one for the central contact of each detector), and three low-resolution, warm-FET position readouts (corresponding to the left two, middle four and right two segments of the detector---see Figure \ref{fig:CloverCartoon}).
Coincidences between the energy and position readouts tell us which segment(s) recorded energy depositions.  All four detectors share a common bias and preamplifier power.

All contacts on the CLOVER were read out with a pair of DGF4C ``Digital Gamma Finder'' digitizer cards from X-Ray Instrument Associates (XIA)\cite{XIA}.  These boards operate under a Computer Automated Measurement and Control (CAMAC) architecture, and sample at 40 MHz (25 ns per sample) with a 14-bit analog to digital converter (ADC).  Oversampling allows an effective resolution in the DGF4C boards of 16 bits.  The data sets described in this paper fall into either ``calibration sets'' (used to calibrate the PSA algorithms) and ``characterization sets'' (used to characterize the efficacy of different event classes under PSA).  For each individual systematic study, the characterization and calibration data sets were taken under identical hardware triggering and event selection conditions.

\section{\label{sec:BGRedPSA}Background Reduction Through Pulse Shape Analysis}
The lack of neutrinos in the final state of \BBz\ means that the only particles to carry away the available kinetic energy (\QBB\ = 2039.04 keV for \nuc{76}{Ge}\cite{QValue}) are the two electrons.  This pair of electrons will have a very limited range in germanium compared to backgrounds near \QBB, such as multiple Compton scatters from higher energy $\gamma$ rays.  Most backgrounds near \QBB\ will tend to produce interaction locations separated by up to several centimeters in germanium, allowing them to spread over a much larger volume and even across several detectors.  \BB\ events on the other hand will tend to stop in the HPGe detector in which they originated, depositing all their kinetic energy with high efficiency.  Therefore, the signature of \BBz\ is a spatially localized ({\it i.e.} single-site) event with a well-defined energy (\QBB), uncorrelated in time with any other event in the data stream.  Pulse shape analysis, like most of the background tagging techniques employed in \nuc{76}{Ge} \BBz\ experiments, exploits the single-site nature of the signal and the multi-site nature of most backgrounds of similar energy.  PSA tends to only be performed only on single-detector events, because any multi-detector event is clearly multi-site.  Nearly all PSA methods rely on digitized waveforms from the detector(s) (some techniques can be implemented only with analog electronics).  The waveforms analyzed here are recorded from the output of the integrating preamplifiers of the detector system.  We refer to such a waveform as a ``charge pulse.''  Many PSA algorithms also analyze the ``current pulse'' of an event rather than the charge pulse.  One can extract the current pulse by simply taking the time derivative of the charge pulse.

Most PSA cut algorithms must first be calibrated on data containing known single and multi-site events, during which the calibration software associates typical waveform characteristics with each class of events (single and multi-site).  For calibration data, we primarily used the full-energy $\gamma$-ray peak at 1588 keV from \nuc{228}{Ac} for our population of multi-site events and the 1592-keV double escape peak (DEP) of the 2614-keV $\gamma$ ray from \nuc{208}{Tl} for our population of single-site events.  DEP events occur when a $\gamma$-ray of sufficiently high energy undergoes electron-positron pair production in a detector, and both 511-keV $\gamma$ rays (created after the positron stops and annihilates) escape the detector without depositing any energy.  DEP events are single-site because the only particles depositing energy in the detector of interest are charged leptons with a short range.  Double escape peaks are described in more detail in \cite[Chapter 12]{Knoll}.  Both the 1588-keV $\gamma$-ray line and 1592-keV DEP are part of the \nuc{232}{Th} decay chain, making it operationally simple to collect both simultaneously.  \nuc{232}{Th} has become something of a standard source in the examination of PSA in HPGe detectors because its decay chain has a relatively high-energy $\gamma$ ray at 2614 keV (giving a double-escape peak at 1592 keV), coupled with a spectrum otherwise free of strong lines above $\sim 1700$ keV.  For some studies, we used the 1771-keV $\gamma$ ray and one of the several DEP peaks available in the spectrum of \nuc{56}{Co}.  Although \nuc{56}{Co} has several high-energy $\gamma$ rays up to 3600 keV, giving DEP lines from 1600 to 2600 keV, spanning \QBB\ for \nuc{76}{Ge}, these $\gamma$-ray lines also lead to a strong Compton continuum in this region, meaning that \nuc{56}{Co} DEP events can be hard to separate from continuum events for the purpose of PSA calibration. \nuc{56}{Co} also has a relatively short half-life (77.27 days), meaning that sources must be replaced rather frequently.  For these reasons, \nuc{232}{Th} tends to be favored for calibration most PSA algorithms.  Throughout this article we will refer to ``DEP events'' and ``$\gamma$-ray events'' as those selected by making energy cuts on these features in either spectrum.  The reverse of the techniques used in \BBz\ searches ({\it i.e.} rejecting single-site events while accepting multi-site events) could have potential applications to searches for rare nuclear decays to excited states, by allowing for the separation of weak $\gamma$-ray lines from a strong continuum.

\subsection{\label{sec:PSAAlgorithms}Evolution of Parametric Pulse Shape Analysis}
PSA was used to varying degrees in both of the previous generation searches for \BBz\ in \nuc{76}{Ge}: the International Germanium Experiment (IGEX)\cite{IGEX99, IGEX02} and the Heidelberg-Moscow experiment (HM) \cite{HM01a, HM01b, HM02, HM04, HMNIM}.  PSA in the HM experiment is discussed in Reference \cite[Section 4.2]{HMNIM}.  The PSA work presented in this article is largely an extension of the PSA work done in support of the IGEX experiment, which is discussed at greater length in References \cite{CraigThesis, Aalseth98, PSASegOrtho}.  Our PSA uses parameters calculated from the charge and current pulses to quantify the single and multi-site nature of events from HPGe detectors.  The way in which an algorithm associates single or multi-site events with numerical values of these parameters represents one of the key differences between parametric PSA algorithms.  This is true of the changes in PSA from References \cite{CraigThesis, Aalseth98} to \cite{PSASegOrtho} to that presented here.

The IGEX experiment used three pulse shape parameters:
\begin{itemize}
\item pulse width (defined as the time it took for the charge pulse to rise from 10\% to 90\% of its full amplitude),
\item front-back asymmetry (defined as the difference in the area of the first and second halves of the current pulse, normalized by the total area of the current pulse), and
\item normalized moment (defined essentially as the moment of inertia of the current pulse, were it treated as a mass distribution---see Reference \cite{PSASegOrtho} for the exact form).
\end{itemize}
The IGEX PSA algorithm was calibrated by creating a three-dimensional histogram of these parameters from a set of DEP events.  The fraction of the total number of calibration events in each bin was then calculated.  The bins of that histogram were rank ordered by that fraction, and stepping down the list, the cumulative fraction of the total number of events in the list was calculated for each element (so that the first element would have only the fraction of events in the most heavily-populated bin and the last element would have all of them).  A cut was then made at an element on this list so that some nominal fraction (80\% in most cases) of calibration events were preserved.  This defined the signal space for the cut.  When production data was later analyzed and its pulse shape parameters calculated, the analysis software checked which bin in the three-dimensional PSA parameter histogram each pulse occupied.  If that bin was in the signal space, it passed the cut.

This method was modified for the preparation of Reference \cite{PSASegOrtho}.  The detectors in the CLOVER are much smaller than those used in IGEX ($\lesssim 0.75$ kg versus $\gtrsim 2$ kg).  As a result, there as little difference in drift times from different regions of the individual CLOVER detectors, so the pulse width was largely degenerate between known single and multi-site events.  Reference \cite{PSASegOrtho} therefore used only the second two parameters listed above.  The calibration software created two-dimensional histograms for both $\gamma$ rays and DEPs from a 25\%-subset of the characterization data.  A fit was then performed to the parameter space histogram for each detector using a complicated analytical model (a sum of many two-dimensional Gaussian and Lorentzian functions).  Both fit functions were then normalized so that their integral over the parameter space was one, allowing for their treatment as probability density functions (PDFs).  When the full characterization data set was analyzed, the analysis software calculated the parameters for each pulse, and the PDF (fit function) for that detector was evaluated at those parameter values.  If the PDF from the DEP ($\gamma$ ray) data had the higher value, then the event in question was classified as single-(multi-)site and passed (failed) the PSA cut.

The PSA algorithm from Reference \cite{PSASegOrtho} was further modified for this work.  Here, we calibrated the PSA algorithm with a data set completely independent of the characterization data (we did quantify the bias resulting from different amounts of overlap in between the calibration and characterization data---see Section \ref{sec:CalibrationData} of this article for more detail).  We also did away with the complicated fit described above.  Instead, we simply normalized the PSA parameter histograms for DEP and $\gamma$-ray events from each detector so that their integral was one, treating them as a PDF.  The analysis of characterization data was then performed by checking the value of the PDF histograms corresponding to the pulse shape parameter values for each characterization event.

\begin{figure}[htbp]
\begin{center}
\includegraphics[angle=90, width=\textwidth]{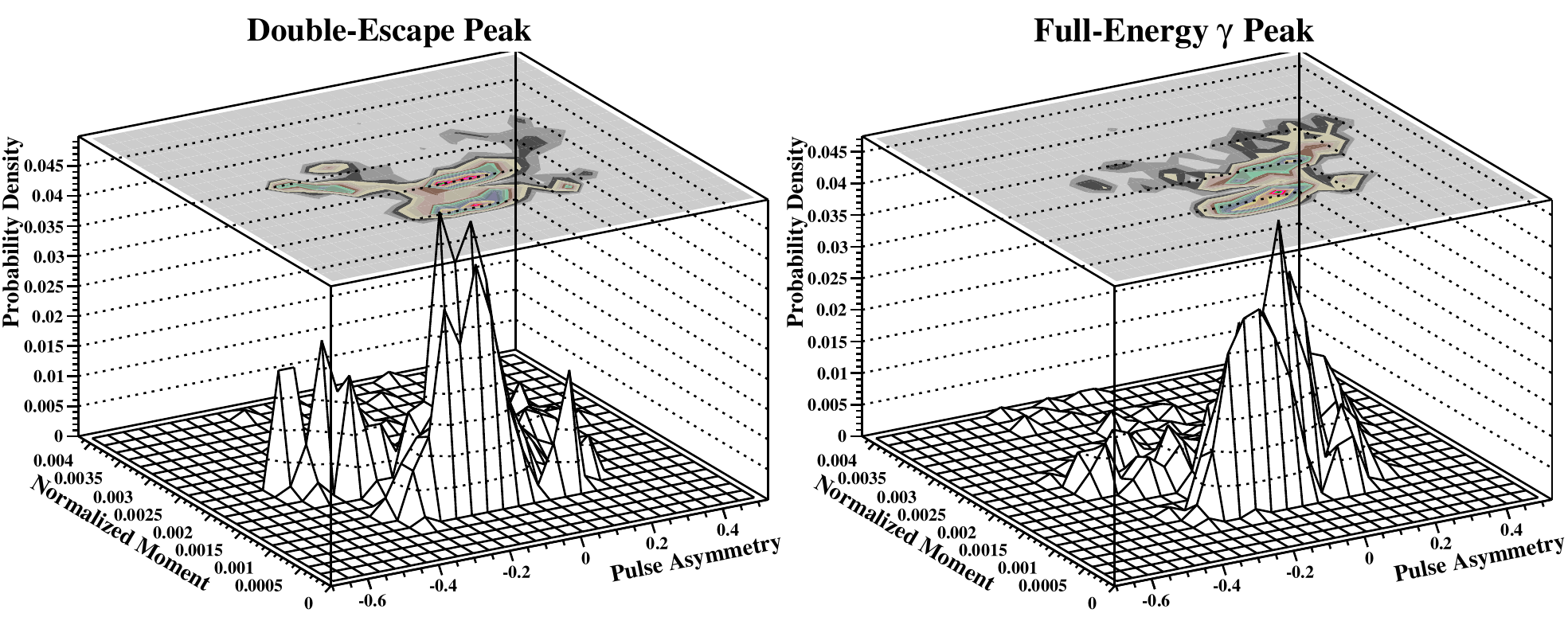}
\caption{\label{fig:PSAKeyComp}A zoomed-in view of the PDF histograms from \nuc{232}{Th} data, populated with 1592-keV DEP (left) and 1588-keV $\gamma$-ray (right) events.  The contour plots are also shown at top of both plots.}
\end{center}
\end{figure}

Because a CLOVER detector is actually a four-detector array, we could select calibration DEPs in two different ways: one in which we made a simple energy cut in the single-detector spectra, and another in which we looked for two-detector events with the DEP energy in one detector, and one of the two $e^{+}$ annihilation $\gamma$ rays in another.  The coincidence tag severely limits the number of DEPs available for calibration but also provides a very clean population.  Most of the results presented in this article were made with single-detector, energy cut-selected DEPs.  PSA calibration $\gamma$-rays were always selected with a single detector cut.  When both analyses are presented, the PDF constructed with non-coincidence tagged DEPs is denoted with a subscript ``NC'' (for ``No Coincidence'').  Figure \ref{fig:PSAKeyComp} shows surface and contour plots of the PDF histograms for single-detector DEP and $\gamma$-ray events.
We could have also tagged DEPs by demanding three-detector events: DEP in one detector, plus both $e^{+}$ annihilation $\gamma$ rays in other detectors.  This would have provided an even cleaner population of calibration DEPs, but the efficiency for capturing such events in our array is impractically low for this study.

To evaluate the efficacy of different implementations of PSA in this article, we examine the ratio of the number of events which pass the PSA cut to the number of events on which the cut is performed, defining this as the survival probability.  Survival probabilities were calculated for three different event classes: the DEP and $\gamma$-ray events discussed above and Compton continuum events near \QBB\ for \nuc{76}{Ge}.  Understanding the survival probability of Compton continuum events is important because they represent the most likely backgrounds to experiments like \mj.  For continuum events, calculating the number of events in the pre and post-PSA cut spectra is simply an exercise in counting the number of events in an energy region.  In this study, the region was either 2.0--2.08 MeV for the \nuc{232}{Th} runs ($\sim 40$ keV on either side of \QBB), or 2.04--2.08 MeV for the \nuc{56}{Co} source runs.  The narrower energy region allowed the analysis to avoid the $\gamma$-ray lines at 2015 and 2035 keV in the \nuc{56}{Co} spectrum.  The statistical uncertainties are then just those arising from Poisson fluctuations.  For peaks on top of continua, the situation is more complicated, because we need to separate out the strength of the peak from that of the continuum on which it rests.  We do so by performing a standard $\chi^{2}$ fit in the ROOT framework \cite{root}.  In this case, the fit model is a flat background plus a Gaussian peak (or two, if multiple peaks reside near each other---this is the case for the 1588-keV $\gamma$-ray peak and 1592-keV DEP).  To extract a peak's strength, we calculate its area from the fit parameters.  The area uncertainty comes from that for the fit parameters and the expression for the area of a Gaussian using standard error propagation techniques.  We can then take the ratio of these pre and post-PSA cut peak areas to calculate the survival probability for that cut.  We again propagate uncertainty through the expression to obtain the survival probability uncertainty.  Throughout this article, this is referred to as the ``fit uncertainty.''  All uncertainties presented are one-sigma, rather than another confidence interval.

\subsection{\label{sec:NomPSAEff}Nominal Pulse Shape Analysis Efficacy and Comparison to Previous Work}
In Table \ref{tab:PSASegEff}, we present nominal values for the survival probability of pulse shape analysis cuts for $\gamma$-ray, DEP, and continuum events along with total uncertainties (fit/statistical plus systematic).  If the PSA algorithm worked perfectly and the different event classes had no overlap in the range of their pulse shape parameter distributions, the survival probability for DEP ($\gamma$-ray) events would have a survival probability of 100\% (0\%).
\begin{table}[h]
\begin{center}
\caption{\label{tab:PSASegEff} PSA Survival probabilities for different processes over a range of energy including systematic uncertainties detailed in Section \ref{sec:SysUnc}.}
\begin{tabular}{c|c|c}
\hline\hline
Process                    &Energy (MeV) &PSA Survival Probability (\%)\\
\hline
\nuc{228}{Ac} $\gamma$ ray &$1.588$      &$29.0 \pm 1.9$\\
\nuc{208}{Tl} DEP          &$1.592$      &$65.6 \pm 2.5$\\
Continuum                  &$2.0 - 2.08$ &$45.4 \pm 0.7$\\
\hline\hline
\end{tabular}
\end{center}
\end{table}

For comparison, Reference \cite{PSASegOrtho} found PSA performance for DEP, continuum and $\gamma$-ray line events to be $75 \pm 2.0 \pm 2.1$\%, $43 \pm 0.9 \pm 3.0$\%, and $20 \pm 0.5 \pm 1.0$\%, respectively.  The first uncertainty is statistical, and the second is systematic.  Statistical uncertainties in Reference \cite{PSASegOrtho} were calculated assuming ``root-N'' Poisson fluctuations of the fit-peak area in each spectrum.  This makes them somewhat smaller than the fit uncertainties presented in this article.  In Reference \cite{PSASegOrtho}, systematic uncertainties were also discussed in a rather narrowly-defined way.  They came only from differences in the survival probabilities between data sets where the source was placed in different locations around the CLOVER cryostat.  We estimated the magnitudes of uncertainties in PSA cut survival probabilities as a function of many systematic effects using the CLOVER.  These were typically variations of the method used to implement the PSA cuts and are discussed at length in Section \ref{sec:SysUnc}.

It is clear that Reference \cite{PSASegOrtho} and the work presented in this article show a modest discrepancy in PSA performance. As stated above, the calibration data used for the PSA in Reference \cite{PSASegOrtho} was a 25\%-subset of the characterization data.  This lead to a bias toward stronger separation of DEP and $\gamma$-ray events.  Furthermore, the complicated fit to the pulse shape parameter PDF in Reference \cite{PSASegOrtho} exacerbated the calibration overlap bias (which is discussed at greater length in Section \ref{sec:CalibrationData} of this article).  The large number of parameters in the PDF fit in Reference \cite{PSASegOrtho} lead to artificial structure in the resulting fit functions.  When we removed this overlap between the calibration and characterization data sets and re-ran the fit-based PSA during the preparation of this article, the survival probabilities come much closer to those presented in Table \ref{tab:PSASegEff} (24.3\% for $\gamma$-rays and 68.6\% for DEP events).  Note that this bias associated with the DEP and $\gamma$-ray events is not present for the Compton continuum data, which was not used for calibration the PSA in either algorithm.  The continuum survival probability in this report and Reference \cite{PSASegOrtho} agree quite well, which is expected as the biases discussed above would only affect those data sets for which events within the calibration and characterization data sets are correlated.  We do not present comparisons between this work and PSA from the IGEX or HM collaborations because the detectors used in those experiments were very different than those used to prepare this article and Reference \cite{PSASegOrtho}.

\section{\label{sec:SysUnc}Systematic Uncertainties in Pulse Shape Analysis}
We now present our detailed investigation of systematic uncertainties in the survival probabilities of different event classes under our PSA technique.  This study can also likely serve as an explicit list of effects that should be investigated for any novel PSA technique.  Many of these are effects which can be completely removed with the careful control of run conditions during the collection of calibration, characterization and production data ({\it i.e.} hardware triggering conditions).  Some however, are true systematic uncertainties that cannot be removed operationally and can only be quantified (such as different calibration data sets or source location).

\subsection{PSA Parameter Histogram Resolution and Calibration Set Size}\label{sec:PSABins}
\begin{figure}[htbp]
\begin{center}
\includegraphics[angle=00, width=15cm]{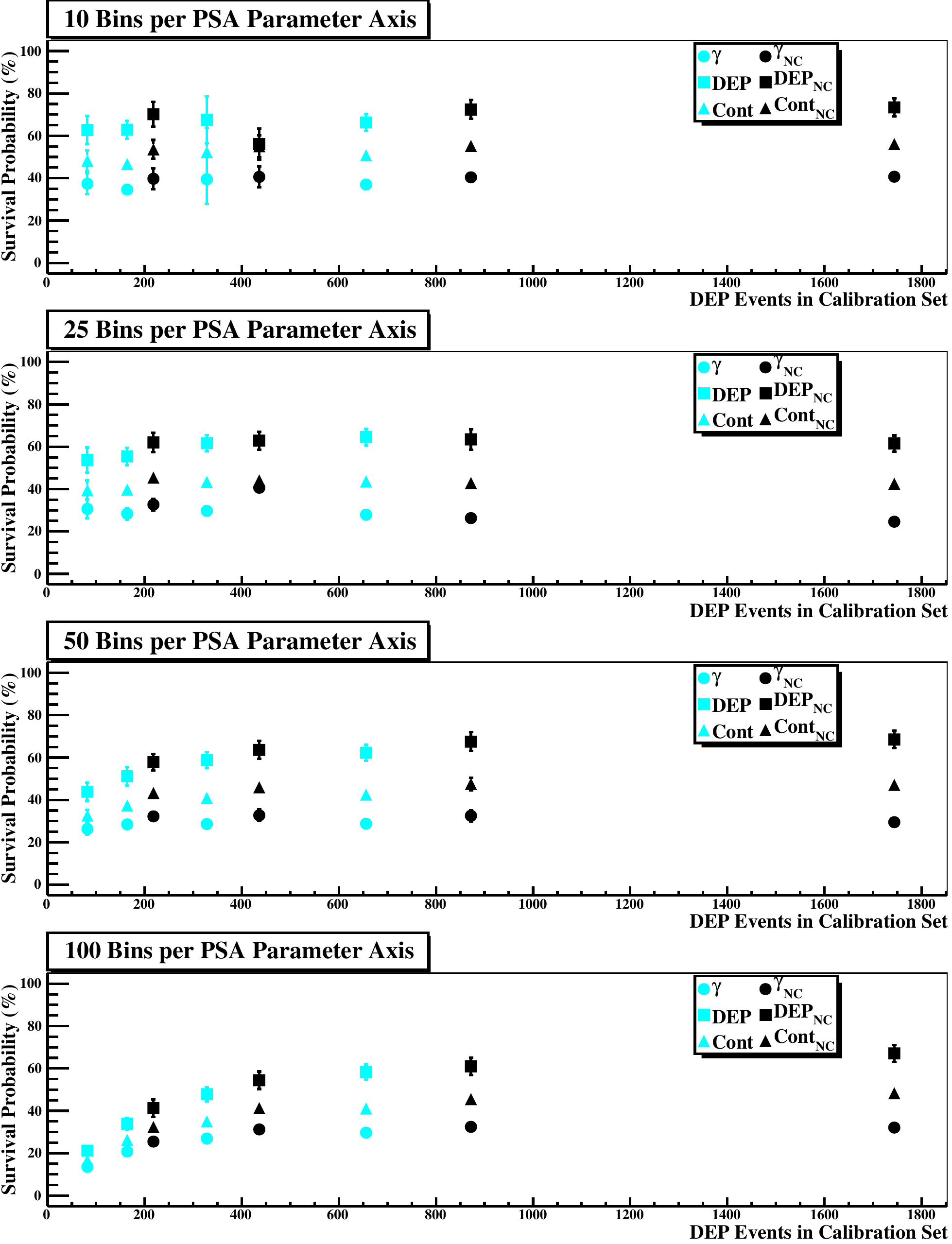}
\caption{\label{fig:CalibrationSizeSummary}PSA efficacies as a function of the number of double-escape peak events per detector in the calibration data set for 10, 25, 50 and 100 bins per PSA parameter axis.  There are multiple data points for the small calibration set sizes because we constructed multiple calibration data sets.  See the text for more detail.}
\end{center}
\end{figure}

The PDF histograms have a finite number of bins in each axis.  We therefore examined how the binning of these axes affects both PSA performance and the required size of the calibration data set.  To this end, we construct a series of plots showing PSA performance as a function of calibration set size for four PDF histogram axis binnings.  Figure \ref{fig:CalibrationSizeSummary} shows this performance for: 10, 25, 50 and 100 bins per PSA parameter histogram axis.  The range of the axes for each PDF histogram was held constant, so we are adjusting the resolution of the parameter space in each PDF.  The calibration set size is characterized by the number of DEP events as they enter the data stream at a much lower rate than $\gamma$-ray events.

First, the figures show that as binning in the PSA parameter histograms becomes finer, the survival probabilities for each event class require larger calibration sets to reach their asymptotic values.  Finer binning generally improves PSA performance, but there is a point of diminishing returns around 50 bins per axis.  That is, increasing the number of bins from 10 to 25 to 50 improves separation of DEP and $\gamma$-ray events, but increasing from 50 to 100 bins per axis does not, even with large calibration sets.  Finer binning in PSA parameter space allows for the resolution of finer structure in the signal and background PDF histograms, however, once that structure is captured, increasing resolution does not seem to improve the algorithm.  We therefore chose the optimal conditions for our ensuing PSA studies ({\it i.e.} the binning that provides the best performance with the lowest number of calibration events) of 50 bins per axis for all other studies presented.  The asymptotically best DEP/$\gamma$-ray separation is reached when there are $\gtrsim 500$ DEP events per detector in the calibration data.  This corresponds to between 2.8 and 5.6 million (11 million) total events with our experimental setup without (with) demanding the coincidence tag.

\subsection{Independence of Calibration and Characterization Data Sets}\label{sec:CalibrationData}
As discussed above, it is important that the PSA calibration data be isolated from the data set used to characterize the efficacy of the cuts. If the data sets overlap, the cuts will appear more effective than they actually are because fluctuations in the pulse shape parameters for the different event classes will get frozen into the PDF histograms. This effect is most easily mitigated by completely separating calibration and characterization data.

\begin{figure}[htbp]
\begin{center}
\includegraphics[angle=90, width=\textwidth]{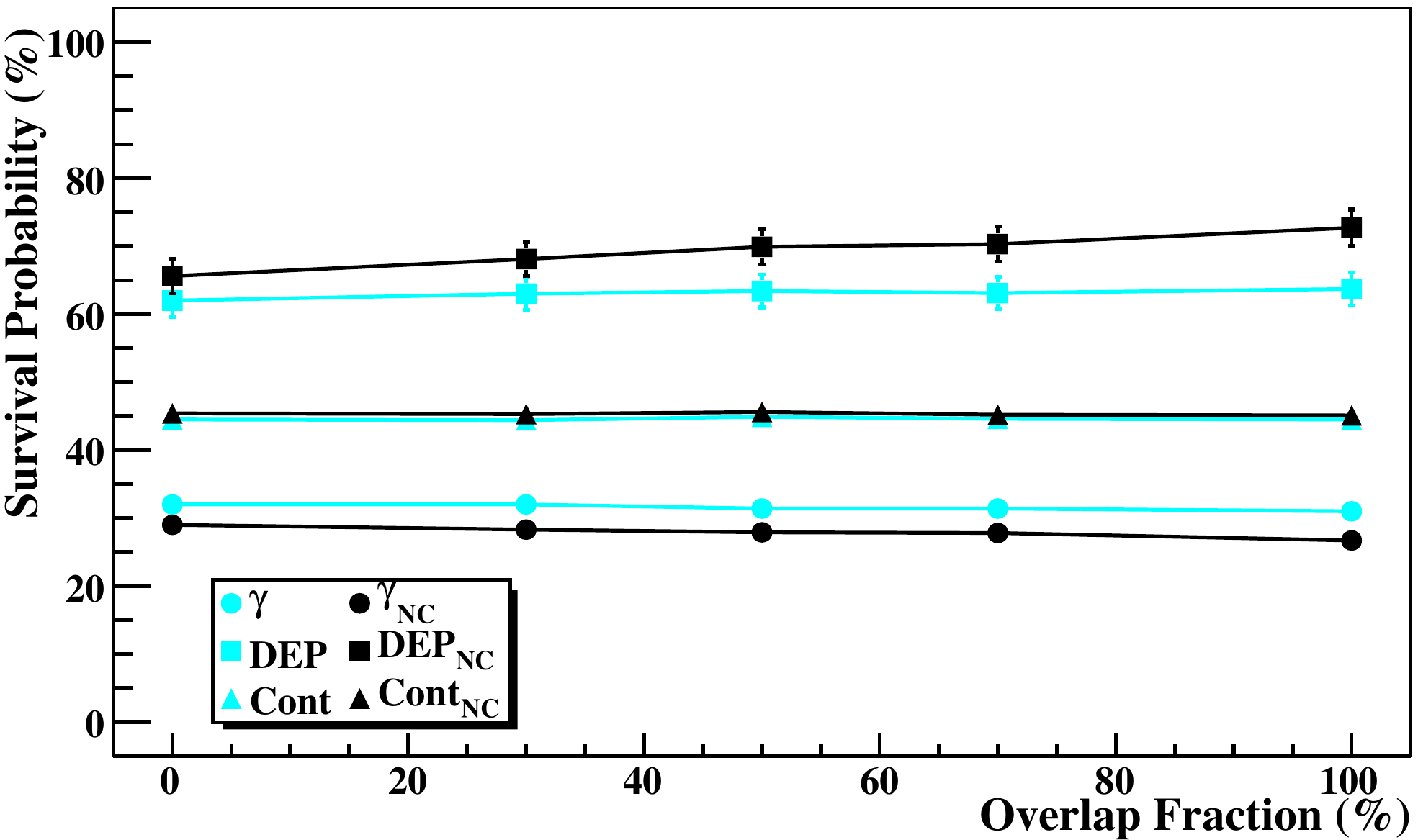}
\caption{\label{fig:OLapStudyThAnConstTrain}PSA survival probabilities as a function of the fraction of the characterization data in common with the calibration data.}
\end{center}
\end{figure}

To quantify the bias in PSA performance due to overlapping calibration and characterization data sets, we used PDF histograms constructed from a single calibration set to analyze several characterization data sets of similar size (there were ten data files containing $7 \times 10^{5}$ events each).  The overlap fraction is set by varying numbers of files in the characterization set in common with the calibration set and calculated by the ratio of the total size of the data files (in megabytes) in both the calibration and characterization data sets to the size of the characterization data set.  We show the results of this study in Figure \ref{fig:OLapStudyThAnConstTrain}.

The figure shows that as the overlap fraction increases, the separation of DEP and $\gamma$-ray events increases for PSA trained on non-tagged DEPs.  For comparison, we also included survival probabilities for PSA  trained on coincidence-tagged DEPs.  The term ``overlap fraction'' is something of a misnomer for the coincidence tagged DEPs because the calibration DEPs are all two-detector events and the characterization DEPs are all single-detector events (all calibration $\gamma$-rays were single detector events regardless of how calibration DEPs were tagged), but we use the above-stated ratio of file sizes to calculate the overlap fraction for comparison.
Figure \ref{fig:OLapStudyThAnConstTrain} indicates that DEP and $\gamma$-ray events for the non-tagged cuts are most strongly affected by sample bias, changing favorably 7.1\% and 2.3\%, respectively.  As expected, the affect is essentially eliminated in the tagged cuts because the coincidence-tagged calibration DEPs never actually enter the characterization data stream.  Finally, we see that the continuum survival probabilities are essentially independent of overlap fraction for the same reason: we do not actually train on continuum events, so there are none in the calibration data regardless of the overlap with the characterization data.

\subsection{Event rate and Pulse Pileup}\label{sec:RateAndPSA}
We now investigate the dependence of PSA performance on count rate in the CLOVER.  Charge pulses from HPGe detectors have a short (typically $< 1\ \mu$s) rise time, containing all of the information about energy deposition.  In detectors with resistive feedback preamplifiers (such as the CLOVER), this is followed by a much longer fall time characterized by the preamplifier's RC time constant (typically $\sim 50\ \mu$s).  This fall time can be removed using a ``Moving Window Deconvolution'' \cite{MVD}, but no such technique was employed in this work.  Therefore, for our PSA to be maximally effective, the event rate had to be low enough that pulses from energy depositions in the detector are not deformed by the tail of the preceding pulse.  If we assume an average time between pulses of ten to twenty preamplifier time constants, we estimate a maximum rate of $\sim 1 - 2$ kHz per detector.  The pulse shape data presented in this article were taken over a range of rates from $\sim 40$ Hz up to $\sim 900$ Hz.  Even at 2 kHz, (over twice our highest data rate), the fraction of the integral of a pulse overlapping another following by ten time constants (500 $\mu$s) should only be at the level of $\sim 5 \times 10^{-5}$.  For reference the single-detector rate expected in \mj\ for production data is something like one count per hour, and no more than a few hundred Hz per detector for calibration.  The principle source of dead time of our data acquisition (DAQ) system is actually communication across the backplane of the CAMAC crate.  This is a much larger problem than pulse pile-up at rates approaching 1 kHz.  The live time in the CAMAC DAQ system drops below 90\% at around $80-90$ Hz, drops below 80\% by $\sim 200$ Hz and by 1 kHz has dropped to $\sim 40$\%.  Dead time from backplane communication should have little or no effect on the discrimination power of PSA.  Nevertheless, we looked for an effect.

We adjusted the input count rate by placing different amounts of a \nuc{232}{Th} source two inches from the front face of the CLOVER within the lead shield.  Event rates ranged from $\sim 40$ to $\sim 500$ Hz.  At each rate, we collected 10.5 million events for each of the three highest rates (98, 143 and 531 Hz), corresponding to $\sim 1500$ DEPs in the single-detector spectrum.  The data set at the lowest rate (39 Hz) is twice as large corresponding to $\sim 3000$ DEP events.  We therefore treat two halves of the low-rate data set independently.  We trained the PSA cuts on the first two thirds of each data set and characterized their efficacy using the remainder.  PSA results are reported in Table \ref{tab:PSARateDepResults}.
\begin{table}[h]
\begin{center}
\caption{\label{tab:PSARateDepResults} PSA Efficacies at different event rates.  Survival probability uncertainties are just the error from the fit.  Event rate uncertainties are the standard deviation of the rate reported by the DAQ software from run to run.}
\begin{tabular}{c|c|c|c}
\hline\hline
Rate (Hz)        &DEP (\%)       &$\gamma$-ray (\%) &Continuum (\%)\\
\hline
$530.8 \pm 39.1$ &$63.2 \pm 4.5$ &$31.1 \pm 1.9$    &$46.8 \pm 1.1$\\
$143.1 \pm 0.1$  &$66.9 \pm 3.6$ &$29.2 \pm 1.3$    &$46.4 \pm 0.9$\\
$97.6 \pm 0.3$   &$66.9 \pm 3.8$ &$33.1 \pm 1.5$    &$48.2 \pm 0.9$\\
$38.8 \pm 0.1$   &$64.8 \pm 4.2$ &$35.9 \pm 1.7$    &$45.9 \pm 0.9$\\
$38.8 \pm 0.1$   &$68.1 \pm 4.0$ &$34.0 \pm 1.7$    &$46.2 \pm 0.9$\\
\hline\hline
\end{tabular}
\end{center}
\end{table}

As expected, we find that the PSA survival probabilities depend quite weakly on the event rate.  We do see that for the 531 Hz data, DEP survival drops slightly but not by a statistically significant amount.  Still, because DAQ dead time from the CAMAC backplane is a problem at a considerably lower rate, we held the event rate in the rest of our studies down to below $100-150$ Hz.  Pulse shape distortion from pile-up could be a concern, however if a more modern DAQ system allowed for higher rates or in the case of two-step decays through states with $\sim \mu$s lifetimes.  In this case, more sophisticated pulse processing would be required to remove the preamplifier's RC time constant.

\subsection{Different Calibration Sets}\label{sec:DifferentCalibrationSets}
Throughout the course of any experiment using PSA, experimenters will no doubt take many calibration data sets (assuming of course, that they use algorithms that require calibration data).  Periodic calibration will also be necessary to check the stability of the detector array over time.  We considered possible uncertainty in pulse shape analysis performance from calibration on different data sets taken under ostensibly the same conditions.

The data sets described in Section \ref{sec:RateAndPSA} recorded at the three lowest rates (39, 98 and 143 Hz), were split into six equal parts, labeled A--F, providing 5 calibration data sets (A--E) and 1 characterization data set (F).  There were a total of 60 700,000-event data files, so each of the six parts contained data files with roughly equal size taken at each event rate.  We generated independent PDF histograms using each of the first five parts (A--E) and used those five PDF histograms to analyze the sixth data set.  We compared the spread of the 5 deduced survival probabilities with their variance.  If the spread is comparable to or smaller than the fit uncertainties  for the individual results, then the fit errors place an upper limit on the calibration set uncertainty.  If the spread is larger than the fit uncertainties, then we can use the difference between the spread and the fit uncertainty to estimate the additional systematic uncertainty arising from fluctuations in the calibration data.
\begin{table}[h]
\begin{center}
\caption{\label{tab:PSASixPartResults} PSA performance  on data set F, trained on data sets A through E.  Uncertainties listed with the survival probabilities come from the peak fit.  The row ``Mean and SD'' reports the mean and standard deviation of the previous five rows.  The row ``$\sigma_{SD} - \overline{\sigma}_{Fit}$'' is the difference between the standard deviation and the average fit uncertainty.  See the text for more detail.}
\begin{tabular}{c|c|c|c}
\hline\hline
Calibration                                &\multicolumn{3}{c}{Survival Probabilities from Analyzed Set}\\
Data Set                                &DEP (\%)       &$\gamma$-ray (\%) &Continuum (\%)\\
\hline
A                                       &$66.3 \pm 2.7$ &$33.5 \pm 1.1$ &$46.6 \pm 0.6$\\
B                                       &$65.8 \pm 2.7$ &$33.9 \pm 1.1$ &$46.9 \pm 0.6$\\
C                                       &$66.0 \pm 2.7$ &$33.8 \pm 1.1$ &$47.3 \pm 0.6$\\
D                                       &$67.2 \pm 2.7$ &$35.9 \pm 1.1$ &$48.2 \pm 0.6$\\
E                                       &$65.7 \pm 2.7$ &$32.6 \pm 1.1$ &$45.9 \pm 0.6$\\
\hline
Mean and SD                             &$66.2 \pm 0.6$ &$33.9 \pm 1.2$ &$47.0 \pm 0.9$\\
$\sigma_{SD} - \overline{\sigma}_{Fit}$ &$-2.1$         &$+0.1$         &$+0.3$\\
\hline\hline
\end{tabular}
\end{center}
\end{table}

The last row of Table \ref{tab:PSASixPartResults} shows us that the spread for DEP survival probabilities is markedly less than the average fit uncertainty.  For $\gamma$-ray and continuum events, it is only slightly larger, with an excess of 0.1\% and 0.3\% respectively.  It is also worth noting that the spread in these values seem to be driven by calibration set D which has consistently higher survival probabilities for all event classes.  We include these results as is, but it is probable that they are upper limits on this uncertainty.

\subsection{Triggering Conditions}\label{sec:TriggeringAndPSA}
The DGF4C DAQ boards have three parameters that configure the triggering conditions: a threshold and two filtering times.  According to Reference \cite{XIA}, one of the trigger times, called the ``Flat Top Time'' affects the digitizers' ability to trigger on slower or faster rising pulses and therefore should most strongly affect the population of single versus multi-site events.  Multi-site events should tend to have slightly longer rise times than single-site events  (though the difference is quite small in the CLOVER).  We therefore investigated the impact of this parameter by examining the data sets taken at 39 Hz for Section \ref{sec:RateAndPSA} (trigger flat top time = 0.2 $\mu$s) and collected another with the same configuration with a trigger flat top of 0.075 $\mu$s.  This spanned the range we had used for this filtering time in our entire experimental program with the CLOVER that still produces proper operation of the detector.  We then analyzed it in the same way as in Section \ref{sec:RateAndPSA} ({\it i.e.} trained on two thirds of the data set and characterized the remaining third).  We treated the two halves of the larger 0.2 $\mu$s data set independently.  We compare the results in Table \ref{tab:56CoTrigCmp}.  To facilitate comparison, we also include the difference between the survival probabilities for data taken with each trigger flat top and each event class.
\begin{table}[h]
\begin{center}
\caption{\label{tab:56CoTrigCmp} PSA cut survival probabilities comparing data taken under different triggering conditions.  The 0.2-$\mu$s data was taken from the 39-Hz event rate data sets in Table \ref{tab:PSARateDepResults}.  The survival probability uncertainties are once again those from the fit peak areas.  The last two lines of the table (labeled ``Difference'') are the differences between the survival probabilities for each event class in the 0.075-$\mu$s data set and each of the 0.2-$\mu$s data sets.  See the text for more detail.}
\begin{tabular}{c|c|c|c}
\hline\hline
Flat Top     &DEP (\%)       &$\gamma$-ray (\%) &Continuum (\%)\\
\hline
$0.075 \mu$s &$64.8 \pm 3.7$ &$29.0 \pm 1.5$    &$40.2 \pm 0.7$\\
$0.2 \mu$s   &$64.8 \pm 4.2$ &$35.9 \pm 1.7$    &$45.9 \pm 0.9$\\
$0.2 \mu$s   &$68.1 \pm 4.0$ &$34.0 \pm 1.7$    &$46.2 \pm 0.9$\\
\hline
Difference   &$ 0.0 \pm 5.6$ &$-6.9 \pm 2.3$    &$-5.7 \pm 1.1$\\
             &$-3.3 \pm 5.4$ &$-5.0 \pm 2.3$    &$-6.0 \pm 1.1$\\
\hline\hline
\end{tabular}
\end{center}
\end{table}

Table \ref{tab:56CoTrigCmp} shows that the survival probabilities for DEP events remained unchanged when we lowered the trigger flat top time.  Those for $\gamma$-ray and continuum events however, are substantially lower with a shorter flat top time.  This leads us conclude that PSA performance can be affected by triggering conditions.  Trigger filter parameters should therefore be tuned for each detector in an experiment to maximize the efficacy of PSA as well as energy resolution and other characteristics.  It is also important that pulse shape analyses of data sets with different triggering conditions be carefully considered.

\subsection{Analyzed Line Energy}\label{sec:LineEnergy}
\begin{figure}[htbp]
\begin{center}
\includegraphics[angle=90, width=\textwidth]{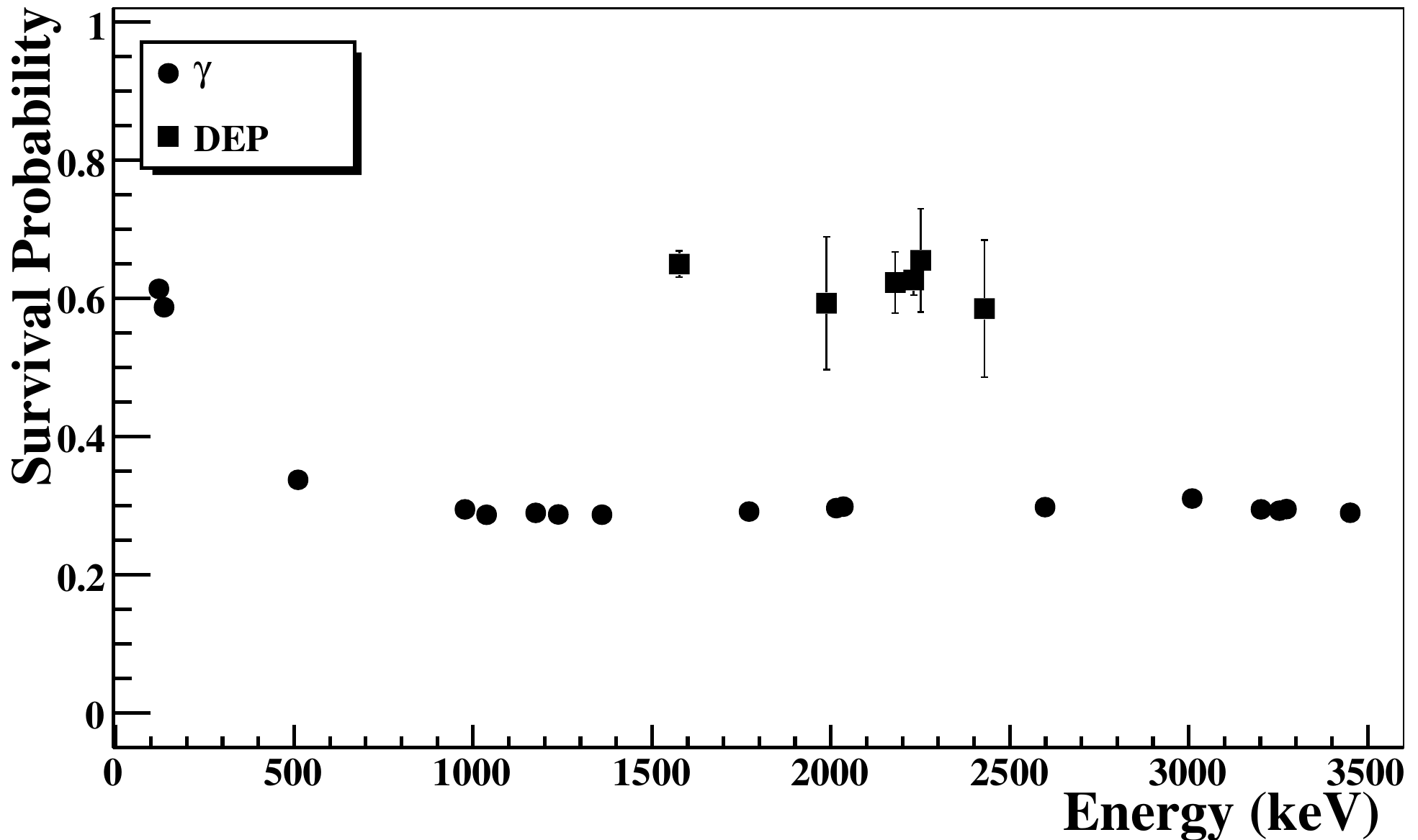}
\caption{\label{fig:SurvProbEDepLine}Line energy dependence of pulse shape analysis cuts for $\gamma$ and double-escape peak events.}
\end{center}
\end{figure}

Figure \ref{fig:SurvProbEDepLine} was constructed by calibration our PSA software on the 1576-keV DEP and 1771-keV $\gamma$-ray lines from \nuc{56}{Co}.  We then plotted the survival probabilities for different lines in the spectrum as a function of energy using these calibration parameters.
Note that the very low-energy $\gamma$-ray lines in Figure \ref{fig:SurvProbEDepLine}, whose interaction is dominated by single photo-absorbtion (a very spatially localized process), have nearly identical behavior under PSA as the DEP lines, as expected.  There is very little variation in the survival probability of $\gamma$-ray lines above $\sim 1$ MeV.  Quantitatively, the PSA survival variation of DEP and $\gamma$-ray events over 1 MeV has a standard deviation of 0.02\% and 0.006\% respectively.  The average fit uncertainty is 0.06\% for DEP and 0.005\% for $\gamma$-ray events.  While the $\gamma$-ray events show a variation slightly in excess of the fit uncertainty, it is exceedingly small at the level of only 0.001\%.  Generally speaking, over the energy range available in \nuc{56}{Co} data (which more than covers the energies of interest for tagging \BBz\ and \BBt\ events), DEPs at different energies behave nearly the same, while $\gamma$-ray events can interact through several different processes \cite{Knoll}.

\subsection{Calibration DEP Energy}\label{sec:TrDEPEnergy}
In principle, calibrating the PSA algorithm on one DEP line should be as good as doing so on any other (over the energy range accessible with a \nuc{56}{Co} source).  We now investigate any variation in pulse shape analysis from the use of different DEP lines to train the cut.  We generated the PDF histogram for each analysis with the 1771-keV $\gamma$-ray line, and one of the five different DEP lines in the \nuc{56}{Co} spectrum from 1576 to 2429 keV.  As with the preceding \nuc{56}{Co} study, DEP and $\gamma$-ray survival probabilities are averaged for all twelve $\gamma$-ray and five DEP lines above 1 MeV, and we quantify the excess uncertainty by looking at the difference between the standard deviation and average fit parameter uncertainty for the survival probabilities.  We also include the survival probability for continuum events with energy from 2.04--2.08 MeV.  We tabulate these results in Table \ref{tab:SurvProbEDepKey}, and find variation of the calibration DEP contributes an uncertainty of: 1.5\% for $\gamma$-ray events and no excess uncertainty for DEP or continuum events.
\begin{table}[h]
\begin{center}
\caption{\label{tab:SurvProbEDepKey}Average PSA survival probabilities for DEP and $\gamma$-ray lines above 1 MeV and continuum events from 2.04--2.08 MeV for PSA trained on different DEP lines.  $\sigma_{SD}$ is the standard deviation of the five survival probabilities in this table and $\overline{\sigma}_{Fit}$ is the average fit uncertainty for each cut.}
\begin{tabular}{c|c|c|c}
\hline\hline
Calibration DEP                              &DEP (\%)           &$\gamma$ (\%)      &Continuum (\%)\\
\hline
1576 keV                                  &$62.2 \pm 2.8$     &$29.4 \pm 0.2$     &$50.0 \pm 0.5$\\
2180 keV                                  &$62.0 \pm 2.8$     &$31.0 \pm 0.2$     &$49.9 \pm 0.5$\\
2231 keV                                  &$63.5 \pm 2.8$     &$30.8 \pm 0.2$     &$50.4 \pm 0.5$\\
2250 keV                                  &$62.9 \pm 2.8$     &$31.8 \pm 0.2$     &$50.3 \pm 0.5$\\
2429 keV                                  &$60.1 \pm 2.8$     &$34.1 \pm 0.2$     &$49.5 \pm 0.5$\\
\hline
$\sigma_{SD}$ - $\overline{\sigma}_{Fit}$ &$1.3 - 2.8 = -1.5$ &$1.7 - 0.2 = +1.5$ &$0.4 - 0.5 = -0.1$\\
\hline\hline
\end{tabular}
\end{center}
\end{table}

\subsection{Source Location Dependence}\label{sec:SourceLocation}
Last, we examine the dependence of PSA performance on the position of the source used to collect the data.  The source was placed above, in front of and to the side of the CLOVER.  The source being placed above and to the side of the CLOVER changes the source's position with respect to the segmentation planes and the crystal axes of the detectors.  Placing the source in front of the CLOVER illuminates the closed end cap parts of each detector as opposed to the parts of the detectors along their cylindrical axes.  We return to generating PSA parameter PDFs using the 1576-and 1771-keV DEP and $\gamma$-ray lines from \nuc{56}{Co}.  Survival probabilities for each event class are again averaged for all DEP and $\gamma$-ray lines above 1 MeV and continuum events from 2.04--2.08 MeV, and we assess the uncertainty from this variation again by looking at the difference between the standard deviation of the survival probabilities and uncertainty from the fit parameters.  We tabulate these results in Table \ref{tab:PSASurvivalsPosVar}.  Once again, we find no variation in the DEP or continuum survival in excess of the fit uncertainties, and a relatively small one for $\gamma$-ray events at 0.8\%.
\begin{table}[h]
\begin{center}
\caption{\label{tab:PSASurvivalsPosVar}Average PSA survival probabilities for DEP events and $\gamma$-ray events above 1 MeV and continuum events from 2.04--2.08 MeV for runs with the source in different positions around the CLOVER.  $\sigma_{SD}$ is the standard deviation of the survival probability and $\overline{\sigma}_{Fit}$ is the average fit uncertainty for each data set.}
\begin{tabular}{c|c|c|c}
\hline\hline
Source Position                           &DEP (\%)           &$\gamma$ (\%)      &Continuum (\%)\\
\hline
Above                                     &$64.0 \pm 6.8$     &$25.3 \pm 0.3$     &$45.7 \pm 0.8$\\
Front                                     &$58.9 \pm 6.1$     &$27.3 \pm 0.3$     &$47.9 \pm 0.9$\\
Side                                      &$59.6 \pm 7.6$     &$27.1 \pm 0.4$     &$48.2 \pm 1.2$\\
\hline
$\sigma_{SD}$ - $\overline{\sigma}_{Fit}$ &$2.8 - 6.8 = -4.0$ &$1.1 - 0.3 = +0.8$ &$0.2 - 1.0 = -0.8$\\
\hline\hline
\end{tabular}
\end{center}
\end{table}

\section{Discussion}\label{sec:Disc}
We now motivate the choice of specific values for the systematic uncertainty for the PSA probabilities.  We summarize the results from Section \ref{sec:SysUnc} now in Table \ref{tab:PSASysSum}.
\begin{table}[h]
%\begin{center}
\caption{\label{tab:PSASysSum} Summary table of systematic effects impacting parametric pulse shape analysis with our CLOVER.  The effects are quoted for PSA cuts trained on non-coincidence-tagged DEPs.  Rows where the effect is marked with a ``*'' superscript can be held to zero operationally for production data taking.}
%\begin{minipage}[htbp]{15cm}
\begin{tabular}{c|l|c|c|c}
\hline\hline
Effect                &Description                                     &$\gamma$ ray (\%)          &DEP (\%)                   &Cont. (\%) \\
\hline
Calibration           &Different numbers of data files in common       &$-2.6 \times \sigma_{Fit}$ &$+2.7 \times \sigma_{Fit}$ &$+0.5 \times \sigma_{Fit}$ \\
overlap$^{*}$         &between calibration and characterization data   &                           &                           & \\
\hline
Event Rate$^{*}$      &Rates $<$ 143 Hz                                &---                        &---                        &---\\
\hline
Different             &Trained on five independent data sets           &0.1                        &---                        &0.3\\
Calib. Sets           &and analyzed same events                        &                           &                           & \\
\hline
Trigger               &Difference between two trigger settings         &3.5                        &---                        &4.7\\
Settings$^{*}$        &minus fit uncertainty                           &                           &                           & \\
\hline
Analyzed              &DEP and $\gamma$ ray events above 1 MeV         &0.001                      &---                        &---\\
Line                  &from \nuc{56}{Co} data set                      &                           &                           & \\
\hline 
Calibration DEP       &Trained on same $\gamma$ ray and different DEPs &1.5                        &---                        &--- \\
\hline
Source Location       &Source above, in front, and to side of CLOVER   &0.8                        &---                        &--- \\
\hline
\multicolumn{2}{r|}{Systematic Sum}                                    &1.7                        &---                        &0.3 \\
\multicolumn{2}{r|}{Fit Parameter Uncertainty}                         &0.9                        &2.5                        &0.6 \\
\multicolumn{2}{r|}{Total Uncertainty}                                 &1.9                        &2.5                        &0.7 \\
\hline
\multicolumn{2}{r|}{\bf Nominal Performance}                           &$29.0 \pm 1.9$             &$65.6 \pm 2.5$             &$45.4 \pm 0.7$ \\
\multicolumn{2}{r|}{\bf Fractional Uncertainty}                        &6.6                        &3.8                        &1.5 \\
\hline\hline
\end{tabular}
%\end{minipage}
%\end{center}
\end{table}
Summing the effects in Table \ref{tab:PSASysSum} in quadrature (thereby neglecting any potential correlations in these effects), we arrive at a total uncertainty of 1.9\% for $\gamma$ rays and 2.5\% for DEPs and 0.7\% for continuum events.  Fractionally (that is, normalized to the survival probability for each event class), this comes to 6.6\%, 3.8\% and 1.5\% respectively.  Reference \cite{Gehman07} sets a goal for total uncertainty budgets required to extract information about the mechanism underlying \BBz\ at roughly 20\% to 50\%, depending on the number of \BBz\ measurements and model space of \BBz\ mechanisms considered.  We can see that the fractional uncertainties in PSA performance will not be a major driver of that total uncertainty budget, but should nevertheless be kept as low as possible to increase confidence in the PSA algorithm.

It is also important to note that the PSA performance demonstrated in this article used standard coaxial HPGe detectors read out with digitizers based on rather old technology.  Current plans for data acquisition for the \mj\ experiment include digitizers which sample at 100 MHz (as opposed to 40 MHz for the DGF4C boards used here).  \mj\ digitizers will also communicate over VME rather than CAMAC architecture, thus avoiding the backplane throughput problems encountered here.  The \mj\ collaboration also plans to field P-type Point Contact (PPC) detectors in its D{\sc emonstrator} phase\cite{MAJORANA_CIPANP}.    These detectors have been superior noise performance and excellent PSA discrimination power\cite{PPC_Luke, PPC_Collar, GERDA_PSA}.  The \mj\ collaboration has already begun purchasing ``Broad-Energy Germanium'' detectors (a particular style of PPC detector) from Canberra for deployment in the D{\sc emonstrator}.  The use of these detectors in lieu of standard semi-coaxial detectors as well as fielding more sophisticated digitization electronics should substantially improve PSA discrimination of the \mj\ experiment over what has been shown here.

\section{Conclusions}\label{sec:Concl}
Much of this and previous work on PSA in germanium-based \BBz\ searches has focused on the notion of reducing backgrounds using PSA, but that is not the primary way in which the next generation of \BBz\ searches will exceed the sensitivity of the previous one.  Enhanced sensitivity for future \BBz\ searches will be achieved primarily by building larger experiments out of cleaner materials, whereas novel analysis techniques will contribute to background reduction at a subdominant level.  It is true however, that all analysis techniques employed in future \BBz\ searches will be examined for systematic uncertainties using a process similar to the one outlined in this article.  This will be particularly important if several positive detections are made in several isotopes and the community undertakes a global \BBz\ analysis like the one in Reference \cite{Gehman07}.  We used DEP and $\gamma$-ray events as sample single and multi-site events, and quoting survival probabilities that are roughly a factor of two to three different between the two classes of events.  It is far more likely that background events near \QBB\ will be Compton continuum events, and PSA (as well as most other background identification techniques) will not work as effectively on them.  The \Tz\ sensitivity for an experiment is inversely proportional to the square root of the background level, given the same level of exposure\cite[Section 4.6]{MJProposal06}.  Reducing a background to 40\% of its initial value (commensurate with the PSA efficacy for continuum events demonstrated here) would correspond to an increase in sensitivity of only around 60\% ($\sqrt{1 / 0.4} = 1.6$).  Even if the combination of many techniques resulted in a factor of 10--20 reduction in background, this would correspond only to a roughly four-fold increase in \Tz\ sensitivity.  While that level of impact on an experimental program is certainly not negligible, it is not the sole reason that the next generation \BBz\ searches should go to great lengths to tag backgrounds.  Another {\it extremely} important reason to identify background events, is to demonstrate the single-site nature of any potential signal seen at or near \QBB.  This would do much to identify that peak as \BBz\ and not some hitherto unidentified background to a much higher degree of certainty.

\section*{Acknowledgments}\label{sec:Ackn}
This work was supported by Los Alamos National Laboratory's Laboratory-Directed Research and Development program and released under report number LA-UR 09-04235. The authors would like to thank Kareem Kazkaz and John Wilkerson for careful reading and constructive comments, as well as the entire \mj\ collaboration for their support in the preparation of this article.

%%\nocite{*}   %% include everything in the .bib file
%%\bibliographystyle{unsrt}
%%\bibliography{PSASystematics_v1.0.bib}

\end{document}